%% file: main.tex
\documentclass[twocolumn]{aastex7}

\usepackage{amsmath}
\usepackage{booktabs}
\usepackage{caption}
\usepackage{graphicx}
\usepackage{longtable}
\usepackage{makecell}
\usepackage{microtype}
\usepackage{multirow}
\usepackage{savesym}
\usepackage{siunitx}
\usepackage{tabularx}

\makeatletter
\newcommand*{\rom}[1]{\footnotesize\expandafter\@slowromancap\romannumeral #1@\normalsize}
\makeatother

\newcommand{\nustar}{\textit{NuSTAR}}

\newcommand{\bettersim}{{\raise.17ex\hbox{$\scriptstyle\sim$}}}
\DeclareSIUnit\angstrom{\text {Å}}
\DeclareSIUnit\erg{erg}
\DeclareSIUnit\gauss{G}
\DeclareSIUnit\year{yr}
\DeclareSIUnit\count{ct}
\DeclareSIUnit\photon{ph}
\begin{document}

\title{A Statistical Survey of Faint Solar X-ray Transients Observed by \textit{NuSTAR}}
\author[0000-0003-2395-9524]{Reed B. Masek}
\affiliation{University of Minnesota, Minneapolis, MN, USA}
\email{masek014@umn.edu}
\author[0000-0001-7092-2703]{Lindsay Glesener}
\affiliation{University of Minnesota, Minneapolis, MN, USA}
\email{glesener@umn.edu}
\author[0000-0002-6872-4406]{Jessie Duncan}
\affiliation{NASA Marshall Space Flight Center, Huntsville, AL, USA}
\email{jessie.m.duncan@nasa.gov}
\author{Kekoa Lasko}
\affiliation{University of Minnesota, Minneapolis, MN, USA}
\email{lasko062@umn.edu}
\author[0000-0003-1652-6835]{Nat{\'a}lia Bajnokov{\'a}}
\affiliation{University of Glasgow, Glasgow, UK}
\email{n.bajnokova.1@research.gla.ac.uk}
\author[0009-0008-2501-9345]{Mary Davenport}
\affiliation{Epic Systems, Madison, WI, USA}
\email{marydavenport@gustavus.edu}
\author[0000-0002-9852-0869]{Marianne Peterson}
\affiliation{University of Minnesota, Minneapolis, MN, USA}
\email{pet00184@umn.edu}
\author{Ian Markano}
\affiliation{University of Minnesota, Minneapolis, MN, USA}
\email{marka092@umn.edu}
\author{Zasha Avery}
\affiliation{University of Minnesota, Minneapolis, MN, USA}
\email{avery217@umn.edu}
\author[0000-0001-8589-3378]{Kristopher Cooper}
\affiliation{University of Minnesota, Minneapolis, MN, USA}
\email{coop0502@umn.edu}
\author[0000-0003-1193-8603]{Iain G. Hannah}
\affiliation{University of Glasgow, Glasgow, UK}
\email{Iain.Hannah@glasgow.ac.uk}
\author[0000-0002-1984-2932]{Brian W. Grefenstette}
\affiliation{California Institute of Technology, Pasadena, CA, USA}
\email{bwgref@srl.caltech.edu}
\author{Stephen M. White}
\affiliation{Air Force Research Laboratory, Albuquerque, NM, USA}
\email{stephen.white.24@us.af.mil}
\author[0000-0001-5685-1283]{Hugh Hudson}
\affiliation{University of Glasgow, Glasgow, UK}
\email{Hugh.Hudson@glasgow.ac.uk}
\author{S{\"a}m  Krucker}
\affiliation{University of Applied Sciences and Arts Northwestern Switzerland, Windisch, Switzerland}
\affiliation{University of California, Berkeley, California, USA}
\email{samuel.krucker@fhnw.ch}
\author[0000-0002-0542-5759]{David M. Smith}
\affiliation{University of California Santa Cruz, California, USA}
\email{dsmith8@ucsc.edu}
\author[0000-0003-2147-9586]{Sarah Paterson}
\affiliation{University of Glasgow, Glasgow, UK}
\email{sarah.Paterson@glasgow.ac.uk}

\begin{abstract}
  In this paper, we use a highly sensitive telescope to characterize solar X-ray transients ranging from microflares in active regions down to weakly energetic brightenings in the quiet Sun.
  X-rays are closely linked to the initial energy release and immediate heating of solar flares, making them invaluable in understanding their driving processes.
  \nustar\ is the first long-term, direct focusing hard X-ray observatory to have observed the Sun, offering a unique opportunity to search for and characterize X-ray events from inside and outside active regions that would be otherwise unobservable.
  We present the first statistical survey of \nustar\ solar observations, characterizing the thermal and possibly nonthermal properties of \num{113} weakly energetic transients down to \SI{e26}{\erg}, making this the first to directly compare events from the quiet Sun to those in active regions.
  Relative to RHESSI microflares, our \nustar\ transients are generally cooler, dimmer, and have slightly steeper spectra.
  Thermal energy content of active region transients appears to be independent of the volume of emitting plasma for transients produced by active regions.
  This is in contrast to those from the quiet corona, which on average have lower energy content, smaller emission volumes, and appear cool but bright rather than hot but dim, suggesting a break in trends from traditional microflares.
  We found no quiet Sun transients with a thermal energy content above \SI{3e27}{\erg}, implying an upper limit on the amount of energy released in plasma above \SI{3}{\mega\kelvin} by quiescent processes.

\end{abstract}

\input{introduction}
\input{transient-identification}
\input{results}
\input{discussion}
\input{conclusion}
\input{acknowledgements}

\newpage
\appendix
\input{detector-gap-examination}
\input{frequency-normalization}
\input{table}

\bibliographystyle{aasjournalv7}
\bibliography{references}

\end{document}

%% file: introduction.tex
\section{Introduction}%
\label{sec:introduction}

The solar corona constitutes the outermost layer of the Sun and is maintained at temperatures exceeding one million degrees, but the observed sustained energy input is insufficient in explaining such high temperatures.
Weakly energetic solar flares and magnetohydrodynamic waves are two mechanisms proposed to constitute this energy to the corona.

Solar flares are discrete, explosive releases of energy sourced by reconnection of stressed coronal magnetic fields.
Flares are observable across all wavelengths and lead to localized heating of the solar atmosphere; however, the large, archetypal flares cannot be responsible for heating the entire solar corona, as they are too infrequent~\citep{Benz-2016}.

\citet{Parker-1988} proposed the nanoflare, defined to be the basic unit of impulsive energy release.
This definition is often exchanged, as it is here, for a more generic alternative where a nanoflare is \textit{any} small, impulsive energy release of $\bettersim$\qty{e24}{\erg}.
Flare occurrence frequency is observed to increase as its thermal energy content decreases, so it is possible that flare-like energy releases at and below nanoflare magnitudes may collectively heat the solar corona~\citep{Hudson-1991}.
Observation of such weak events is challenging, particularly in the X-rays, but there are studies that have claimed to have observed individual nanoflares in the extreme ultraviolet (EUV)~\citep{Purkhart-2022} and radio~\citep{Mondal-2020} or the effects of nanoflare heating in X-rays~\citep{Ishikawa-2017}.
EUV nanoflare analyses have placed constraints on events with energies unobservable in X-rays, but narrowband temperature biases and selection effects inherent to EUV observations can miss the hottest/peak temperatures in events, complicating energetic analysis of the hottest temperatures, and making it difficult for a direct comparison to X-rays~\citep{Aschwanden-2002, Hannah-2011}.
Further, it is unknown whether EUV nanoflares are physically similar to faint X-ray transients or if they are distinct phenomena.
Thus, it is possible that they have separate occurrence frequency distributions~\citep{Hannah-2011}.
We hereafter refer to a ``transient'' as \textit{any} sudden release of energy, regardless of mechanism and energy content, such that nanoflares, microflares, X-class flares, campfires, etc. all refer to phenomena that fall under this umbrella term.

A competing theory to nanoflare heating is wave heating.
\citet{Biermann-1946} originally proposed acoustic waves to be the main driver of chromospheric heating and is a topic still under investgation through the use of simulations \citep{Kuzma-2019}.
Magnetohydrodynamic waves also hold candidacy, particularly for coronal heating, but it is still unclear how the energy is transported to the corona and is dissipated ~\citep{DePontieu-2007,VanDoorsselaere-2020}.
While it is likely both nanoflare and wave heating mechanisms are at work, it is unknown which is more dominant and under what conditions they operate.

Contrary to nanoflares, modern X-ray instrumentation can unambiguously resolve individual microflares, and recent microflare studies imply they share common operating mechanisms with large flares, notably in particle acceleration and heating.
The Reuven Ramaty High Energy Solar Spectroscopic Imager (RHESSI) was an indirect imaging spectrometer that performed X-ray and gamma ray measurements of solar flares~\citep{Lin-2003}.
A statistical study featuring tens of thousands of RHESSI microflares was performed by~\citet{Christe-2008} and~\citet{Hannah-2008}, characterizing their thermal and nonthermal properties relative to those of larger flares.
While RHESSI could detect weak microflares with thermal energies down to \SI{\bettersim e27}{\erg}, it had sensitivity limitations below \SI{\bettersim e28}{\erg}.

The \textit{Nuclear Spectroscopic Telescope ARray} (\nustar) is a direct focusing X-ray imaging spectrometer designed to observe faint astronomical sources, making it uniquely sensitive for solar observations~\citep{Harrison-2013,Grefenstette-2016}.
Previous \nustar\ solar investigations have demonstrated its value in a solar context; \nustar\ can observe transients as faint as GOES sub-A-class and as bright as GOES B-class~\citep{Hannah-2019,Cooper-2021,Duncan-2021}, it is capable of detecting weak nonthermal emission from accelerated electrons~\citep{Glesener-2020,Polito-2023,Cooper-2024}, and it has been shown to constrain the distribution of hot plasma in both quiet Sun (QS) and active region (AR) conditions, setting constraints on flare heating of the solar corona~\citep{Hannah-2016,Wright-2017,Marsh-2017,Marsh-2018,Paterson-2023,Paterson-2024,Duncan-2024}.

Transient events weaker than GOES B-class are understudied in X-rays relative to more energetic flares due to sensitivity limitations of solar-dedicated instrumentation.
This work aims to characterize the spectral properties and occurrence frequency of weakly energetic X-ray transients---including microflares---observed by \nustar, making this the most comprehensive analysis of the faintest-observable solar X-ray transients to-date.
\nustar's sensitivity makes it currently the \textit{only} instrument capable of performing the measurements necessary for this study, as we feature both active region and quiet Sun transients.
The automated feature identification algorithm and spectral modeling is detailed in Section~\ref{sec:transient-identification-and-analysis}.
Results from the survey are presented in Section~\ref{sec:results}, which are then discussed in Section~\ref{sec:discussion}.
Section~\ref{sec:conclusions} states our conclusions for this study.

%% file: transient-identification.tex
\section{Transient Identification and Spectral Analysis}%
\label{sec:transient-identification-and-analysis}

\subsection{Solar Observations with \nustar}

\nustar\ is an Earth-orbiting satellite with an orbital period of about \SI{90}{\min}, during which approximately \SI{60}{\min} is spent in sunlight.
Its orbital trajectory occasionally crosses the South Atlantic Anomaly (SAA) resulting in \SIrange{10}{15}{\minute} of data loss during its transit.
\nustar\ features two quasi-identical hard X-ray (HXR) telescopes, each consisting of its own set of Wolter-I optics focusing onto a focal plane module (FPM) made up of a \numproduct{2 x 2} grid of CdZnTe pixel detectors.
Each FPM (A and B) has a \qtyproduct{12 x 12}{\arcminute} field of view (FOV) of the sky and can measure individual X-rays and their energies.
Due to its direct imaging and low-volume detectors, \nustar's non-solar background is at least four orders of magnitude lower than that of RHESSI, but its limited photon throughput ($\bettersim \SI{400}{\count\per\second}$ per telescope) reduces its dynamic range for bright, low-livetime sources~\citep{Grefenstette-2016}.
For this work, it is generally accepted that \nustar\ observes energies from \SI{2.5}{\kilo\eV} up to $\bettersim$\SI{12}{\kilo\eV}, and high A-class microflares are the brightest \nustar\ can observe without requiring complicated analysis.

Additional observational limitations arise when \nustar\ is viewing the Sun.
There is a loss of absolute pointing knowledge as its forward-facing star-tracking camera is rendered unusable, resulting in up to \SI{\bettersim 1.5}{\arcminute} uncertainty~\citep{Grefenstette-2016,Glesener-2017}.
The lack of the forward-facing camera also causes abrupt, undesirable shifts in the spacecraft's pointing, giving an apparent motion to any sources within the FOV.
\citet{Grefenstette-2016} described and~\citet{Glesener-2017} demonstrated an automated method with which one can reduce the effects of these shifts \textit{if at least one bright, isolated source is persistent within the FOV}.
However, there are some quiet Sun observations for which no source is within \nustar's FOV that could be used a reference to perform such a correction, so for this work a manual approach was taken to accommodate these shifts (described later; Section~\ref{subsubsec:grouping-clusters-in-time}).

\nustar\ has totaled over \num{200} hours of solar data since its first observation of the Sun in 2014.\footnote{For an overview of the \nustar\ solar pointings see \url{https://ianan.github.io/nsovr}}
Our study utilizes all data through the end of 2022, excluding the first solar observation on September 10 \& 11, 2014 which operated as an engineering test.
We also exclude mosaic observations, when the solar disk is divided into a grid of tiles and \nustar\ observes each tile for \SIrange{\bettersim 1}{3}{\min} each (e.g.~\citet{Paterson-2023}).
Mosaics were excluded since the exposures are too short to apply our automated detection.

\subsection{Transient Identification Algorithm}%
\label{subsec:transient-finding-algorithm}

For each orbit of \nustar\ solar data, we apply our systematic, automated approach to look for transient brightenings in each FPM using energies greater than \SI{2.5}{\kilo\eV}.
The core of our algorithm is based on one first used by~\citet{Berghmans-1999} to find EUV brightenings in ARs that was later adopted by~\citet{Kobelski-2014} and~\citet{Kobelski-2014-HiC} for use with soft X-ray data.

\subsubsection{Building Macropixels}

\begin{figure*}[t!]
  \begin{center}
    \includegraphics[width=\linewidth]{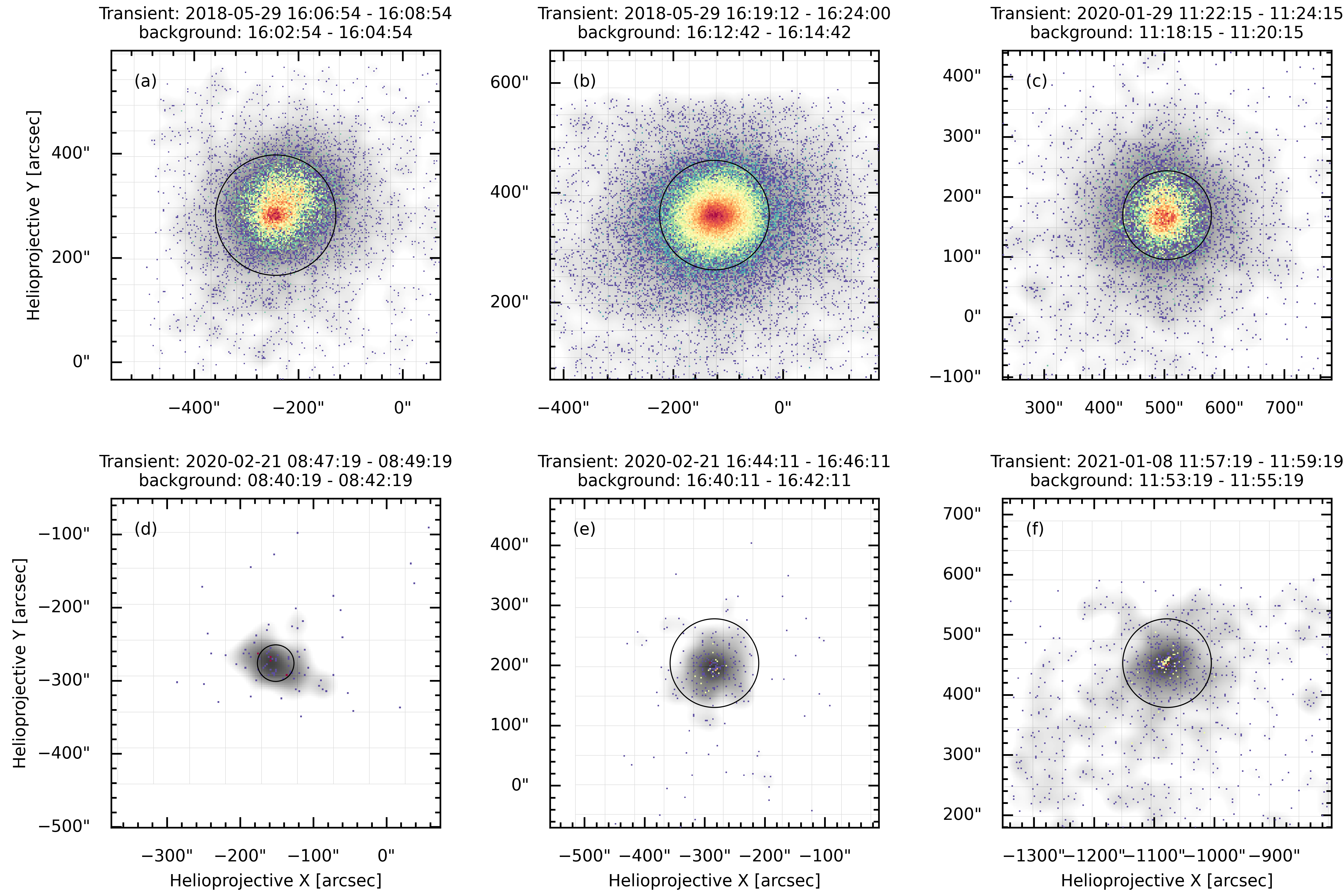}
  \end{center}
  \caption{Examples of identified transients. The top row is a collection of AR transients (with their AR backgrounds), and the bottom showcases QS transients. The rainbow color scale shows counts associated with the transient (more red indicates more counts), the gray, cloudy blur is the Gaussian-smoothed background data co-aligned to the transient data, and the black circle is the chosen spectroscopy region. The grid marks the macropixels.}%
  \label{fig:exposure-map-array}
\end{figure*}

The first step is to spatially bin the image pixels into ``macropixels'' and compute their lightcurves.
``Image pixels'' refer to the post-processed pixels that define the X-ray locations in the sky (e.g.~Figure~\ref{fig:exposure-map-array}), each subtending \SI{\bettersim 2.5}{\arcsecond}.
These differ from the ``detector pixels,'' which refer to the physical pixels of the CdZnTe detectors, each subtending \SI{12.3}{\arcsecond} \citep{Harrison-2013}.
Image pixels attain a higher resolution than detector pixels through interpolation by accounting for charge sharing.
Our specific macropixel bin width ($b=20$) was chosen so that each macropixel has a width of \SI{50}{\arcsecond}, which is close to \nustar's half-power diameter (HPD) of \SI{58}{\arcsecond}.

The motivation for spatially binning the pixels stems from \nustar's relatively coarse spatial resolution (\SI{58}{\arcsecond} HPD), making it often difficult or impossible to resolve loop structures or footpoints asssociated with a transient.
Thus, the binning process does not remove useful spatial information and strengthens the statistics of the signal.
Macropixels are only used for transient identification and are not used in the later analysis, where we return to using the image pixels.
Examples of various transients seen in \nustar\ are shown in Figure~\ref{fig:exposure-map-array}, exhibiting this coarse spatial resolution and how the macropixels compare to the image pixels.

The macropixel lightcurves use \SI{12}{\second} bins and are livetime-corrected as described by~\citet{Bhalerao-2012}.
\nustar\ is capable of measuring the livetime between two photon events to an accuracy of \SI{1}{\micro\second}~\citep{Harrison-2013}.
We compute the mean livetime percentage for each \SI{12}{\second} time bin and compute an effective exposure time by multiplying by the bin width, and the average, livetime-corrected count rate is calculated by dividing the count value by the effective exposure time.

The next few steps follow the general methodology used in the two studies performed by~\citet{Berghmans-1999} and~\citet{Kobelski-2014}.
A moving median (i.e. boxcar) with a window width $w$ is applied to each macropixel lightcurve to calculate the average count and average count rates.
We used a window width of \num{51} frames, equal to \SI{612}{\second}.
A width of about \SI{10}{\min} was selected to function as a background estimation in order to filter out the slowly varying components like active regions, since solar transients (e.g.~microflares) are impulsive on the order of a few minutes~\citep{Christe-2008}.
An odd number (\num{51} instead of a flat \num{50}) was selected for the window width to keep the edges of the time bins aligned with each other.

The residual count rate for each time bin is computed by subtracting the averaged count rate from the count rate to represent the non-flaring background.
The uncertainty of the measured counts is taken to be the square root of the average counts and is then livetime-corrected to obtain uncertainties for the count rates.
If greater than 25\% of the time bins within a boxcar bin had zero counts, the uncertainty for that bin was taken to be \num{4} counts, which is equvalent to roughly two standard deviations according to~\citet{Gehrels-1986}.
Finally, a residual ratio is computed as the residual rate divided by the uncertainty of the average count rate, and this produces a coarse non-flaring background against which we can search for transients.

\subsubsection{Clustering Macropixels in Space}

The next stage of the analysis clusters together macropixels that exhibit a significant signal to construct the spatially-extended X-ray source.
A ``cluster'' is defined as a group of macropixels associated with the same transient, and the spatial properties of individual macropixels are used to associate them with a cluster in a given frame.

The moving median functions as a background estimation, so the residual rate is interpreted as a signal.
Two unitless thresholds are used to determine the significance of a signal: the detection threshold, $q_d=5$, and the connection threshold, $q_c=3$.
The detection threshold sets what value of the residual ratio qualifies as a detected signal, and the connection threshold identifies which neighboring macropixels should be connected to the macropixel with the detected signal.
We selected a high $q_d$ in order to minimize the number of false positives from pointing shifts and shift-related effects (i.e.~transient ``ringing'' seen in lightcurves presumably due to how the mast sways and settles after pointing motion).
A value for $q_c$ lower than $q_d$ was chosen to obtain a more accurate spatial extent since the macropixels on the outer edges of a transient will contain fewer photons due to \nustar's point spread function.

A cluster begins with a macropixel that surpasses $q_d$ in a given time bin, and all of its neighboring macropixels are searched for residual ratios that surpass $q_c$.
All neighbors with a residual ratio greater than or equal to $q_c$ are added to the cluster, and the neighbors of \textit{those} macropixels are also compared against $q_c$.
This is repeated until no additional neighboring macropixels qualify to be added to the cluster, akin to a typical flood fill algorithm.
At this point, the cluster is completed, and the algorithm repeats this process for any remaining, unclustered macropixels surpassing $q_d$ in the given time bin.
This clustering recipe is repeated for each time bin, producing a series of clusters that are next grouped across time.

\subsubsection{Grouping Clusters in Time}%
\label{subsubsec:grouping-clusters-in-time}

With the clusters identified, we reconstruct the brightening event based on the temporal properties of the clusters, where an ``event'' is defined to be a transient candidate and must consist of at least two clusters (i.e. span two frames).
The algorithm begins with a cluster and examines the next time bin for a cluster in a similar spatial location.
A tolerance of one macropixel difference in the spatial positions between clusters in neighboring time bins is allowed to account for possible motion of the plasma or the instrument pointing.
If a cluster in the next time bin is co-spatial, the two clusters are added to the same event, and the algorithm repeatedly adds temporally neighboring clusters for subsequent time bins until they no longer meet the criterion and the event is reconstructed.
This process repeats until all clusters are associated with an event.
Following, each event is then characterized by its macropixel content, time interval, and spatial properties, and products are generated for later use in spectroscopy.
If the duration of an event is less than \SI{2}{\min}, we pad the interval around its midpoint such that the full duration is \SI{2}{\min}.
This is done to ensure proper temporal sampling of the event, which was seen to be otherwise shortened (down to only a few frames, well less than a minute) due to our strict selection of $q_d$ and $q_c$ artifically restricting our time interval.
Our choice of \SI{2}{\min} is rather arbitrary, guided only by the shorter durations observed in RHESSI microflares (\SI{2.2}{\minute};~\citet{Hannah-2008}).

It was previously mentioned that \nustar's pointing can abruptly shift.
These shifts can be detected as candidate transients if there is a persistent source (e.g.~an AR) because the relative motion of the source sweeps across several macropixels, resulting in an apparent brightening.
However, transient candidates of this nature have features that distinguish them from true transients.
Namely, an exposure image of a shift often reveals an elongated, nonlocalized structure compared to true transients.
Additionally, the lightcurves for shifts display very sharp increases/decreases that are likely too rapid (on the order of \SIrange{10}{20}{\second}) for magnetic energy release.
These unique properties distinguish the shifts from the true transients, so each candidate event was manually vetted to remove all shifts from our sample.

\subsubsection{Region Selection}%
\label{subsubsec:region-selection}

Photon data for spectral fitting is extracted from circular regions that are selected in an automated fashion.
We start with a circle centered on the event's brightest macropixel with its radius encompassing all other triggered macropixels in the event.
The region center is then moved to the mean of the \num{20} brightest \textit{image} pixels contained within the original region.
If the region crosses the edge of the FOV, its radius is then iteratively shrunk until it no longer crosses or when it reaches a radius of \SI{40}{\arcsecond} to ensure good sampling of the emission.
Since we do not have absolute pointing information, the FOV is estimated by fitting a minimum bounding box to an exposure of \textit{all} emission recorded within the time interval of the candidate event.
Examples of transients are shown in Figure~\ref{fig:exposure-map-array} with their selected spectral regions.

Our region selection allows the circle to cross the plus sign-shaped gap, referred to as the ``detector gap,'' resulting from the \numproduct{2 x 2} grid construction of the detectors.
Conventionally, solar analyses with \nustar\ avoid the detector gap for region selection, but we opted to remove this constraint in order to increase the number of photons used for spectroscopy and reduce errors introduced by the automated region selection.
Appendix \ref{app:detector-gap-examination} details additional reasoning and a demonstration of the validity of our choice.

\subsection{Spectral Characterization}%
\label{subsec:spectral-characterization}

The XSPEC fitting software, specifically the PyXSPEC Python wrapper, was used to characterize the spectral properties of the found transients~\citep{Arnaud-1996}.
Each transient was independently fit to three models: an isothermal (\texttt{vapec}) model, a double isothermal (\texttt{vapec+vapec}) model, and a thermal with nonthermal (\texttt{vapec+bknpower}) model.
The \texttt{bknpower} model is a broken power law fit in photon space, as XSPEC does not have built-in thick- or thin-target electron models.
The photon index at energies lower than the break energy was held fixed at a value of \num{2} since those energies are predominantly thermal and our spectra are therefore not sensitive to that parameter.
The lower and upper limits reported for all parameter values are the 90\% confidence region obtained from XSPEC's \texttt{error} command (\textit{not} using Monte Carlo chains).
For the transients triggered in both FPMs, the shorter of the two time intervals was used for data extraction, and the data from both FPMs were fit simultaneously to the same model (i.e.~they shared parameters) with a multiplying factor on FPM B to account for systematic differences between the two detectors.
When interpreting the results, the emission measure was multiplied by this factor when it was greater than \num{1}, since this implies that FPM B captured emission that FPM A did not due to where the transient was located within each FPM's FOV.
To accommodate the transients with low photon counts, we utilized C-statistic minimization for all fits.
We also binned all spectra so that each energy bin had at least \num{10} counts.
The models for all transients were bounded by the same parameter limits rather than having catered bounds for each transient to eliminate bias based on the energy content of the transients.
The chosen parameter bounds were informed by prior \nustar\ solar studies and are summarized in Table~\ref{tab:parameter-bounds}.
We assumed static coronal elemental abundances as described in~\citet{Feldman-1992}.

\begin{figure*}[t!]
  \centering
  \includegraphics[width=\linewidth]{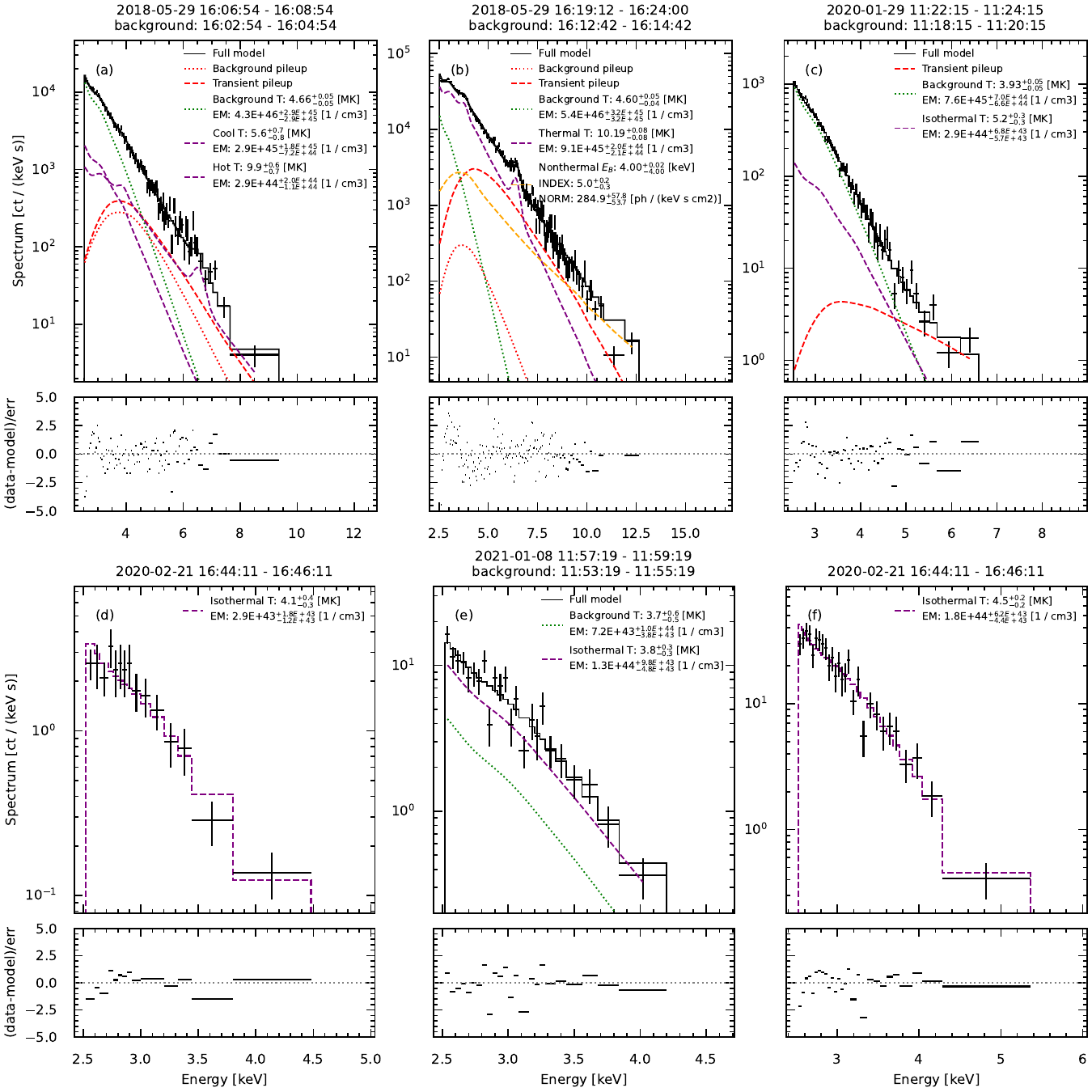}
  \caption{Examples of spectra and models for the transients shown in Figure~\ref{fig:exposure-map-array}. The top row shows AR transients while the bottom row shows QS transients. Only a single FPM is shown in each panel for readability, even if two FPMs were used for fitting. For each transient, the top panel shows the spectra (black crosses with uncertainties) with the fitted model components (solid/dashed/dotted lines) while the bottom panel shows the model residuals normalized by the bin error.}%
  \label{fig:example-spectra}
\end{figure*}

\begin{table}[t!]
  \centering
  \begin{tabular}{l l l}
      \centering
      Quantity & Symbol & Limits \\
      \hline\hline
      Temperature & $T$ & [2.5, 20] \si{\mega\kelvin} \\
      Emission measure & $\text{EM}$ & [\num{1e41}, \num{1e49}] \si{\per\centi\meter\cubed} \\
      Photon index & $\gamma$ & [1, 20] \\
      Break energy & $E_B$ & [4, 20] \si{\kilo\eV} \\
      \hline
      Rate & $\lambda$ & [0.1, 5] \si{\count\per\kilo\eV\second} \\
      Mean & $\mu$ & [2.5, 10] \si{\kilo\eV} \\
      Std. dev. & $\sigma$ & [0.01, 5] \si{\kilo\eV} \\
  \end{tabular}
  \caption{Parameter bounds set for spectroscopy. The parameters above the dividing line are the transient model parameters for the \texttt{vapec} and \texttt{bknpower} components while those below are for the exponentially modified Gaussian used to model pileup.}%
  \label{tab:parameter-bounds}
\end{table}

All photon spectra were made using \nustar\ pixel grades \numrange{0}{4}, which constitute the single pixel and double pixel single-photon events.
Additionally, two optional corrections were added when certain criteria were met: a pileup correction and a gain correction.
These are detailed in Sections~\ref{subsubsec:pileup-correction} and~\ref{subsubsec:gain-correction}.
Examples of fitted spectra are shown in Figure~\ref{fig:example-spectra} for AR and QS transients.

\subsubsection{Background Modeling}%
\label{subsubsec:background-modeling}

Many of our transients were sourced from ARs, necessitating a spectral background correction.
Our condition for requiring a background model was that the selected background spectrum had at least \num{30} counts for each FPM used.
Selecting the background time intervals began by starting with a two minute-long interval beginning four minutes prior to the start of the transient, checking if a previous transient occurred during that time, and selecting the background of the prior transient (if present).
We then validated each background through manual inspection and made minor adjustments to the start and end of the background interval where necessary (e.g.~contaminated by a transient), while always keeping the total duration of the interval at two minutes.
For region selection, we began with the transient region and then re-centered to the mean of the \num{20} brightest pixels during the background interval and shrunk the radius if it crossed the edge of the FOV.
This was done in case a pointing shift occurred between the background and transient intervals.
The background model is an isothermal (\texttt{vapec}) component that was fit \textit{prior} to the transient model and was then held \textit{fixed} while the transient model was fit.
When both FPMs were used, they shared the same background parameters, and a factor was applied to FPM B (exactly as performed for the transient model).
Examples of the background fits are shown by the green dotted lines in Figure~\ref{fig:example-spectra}.
It was found that pileup could also be significant in the background data, so a separate correction to the background was applied independently of the transient pileup correction.

\subsubsection{Pileup Correction}%
\label{subsubsec:pileup-correction}

Pileup is defined as two or more photons incident on the detector such that they are read out as a single photon hitting the detector~\citep{Bhalerao-2012}.
In the case of \nustar's pixel detectors, a photon deposits energy that is distributed among the detector pixels, and the electronics record the incident pixel along with its 8 closest neighbors.
Each recorded photon event is assigned a grade based on the combination of pixels triggered by the absorbed energy, and certain combinations of pixels are impossible---or ``unphysical''---under an assumed charge sharing model describing one photon's energy deposition~\citep{Bhalerao-2012,Bajnokova-2024}.
Pixel grades \numrange{21}{24} include pixels that are only diagonal to each other and are assumed to result from two photons simultaneously incident on the detector.

Our pileup correction follows the methodology in~\citet{Bajnokova-2024}, which is outlined as follows.
Photon spectra made from grades \numrange{21}{24} are fit to an exponentially modified Gaussian.
If both FPMs are used, then each FPM is given its own pileup model since the location of the source within each detector can change the pileup properties in the given detector.
The pileup model normalization is multiplied by a factor of $5/4$ and is then added to the transient model.
This $5/4$ factor was chosen because five pixel grades (\numrange{0}{4}) constitute the transient spectrum and four pixel grades (\numrange{21}{24}) constitute the pileup spectrum, and it is assumed to first order that all piled up events in grades \numrange{0}{4} are equally likely to be a result of two piled up photons.
\citet{Bajnokova-2024} demonstrated that this model and the associated assumptions are adequate by finding excellent agreement for a microflare co-observed between \nustar\ (with piled up spectra) and the Spectrometer/Telescope for Imaging X-rays (STIX) on Solar Orbiter.

A pileup correction is only necessary for events with a high rate of incident photons.
We require this correction when the total flux of photons \SI{>3}{\kilo\eV} for grades \numrange{21}{24} is greater than 1\% of the total flux for grades \numrange{0}{4}.
This value of 1\% was empirically determined through examining a few transients by eye.
When a pileup model is added, it is fit simultaneously to the transient or background model.
Examples of pileup fits are shown by the red dashed (transient) and dotted (background) lines in the top row of Figure~\ref{fig:example-spectra}.

\subsubsection{Gain Correction}%
\label{subsubsec:gain-correction}

Prior solar studies with \nustar\ have noted a shift in the energy gain of the detectors under low livetime conditions that causes an apparent decrease in the observed photon energies~\citep{Duncan-2021,Polito-2023,Duncan-2024,Bajnokova-2024}.
A correction to the detector gain slope is deemed common practice in \nustar\ solar spectroscopy, and we enabled a linear gain correction as prescribed by~\citet{Duncan-2021} for transients with an average livetime less than 1\%.
The gain offset is not fit and is held fixed at a value of \num{0}.
When the slope correction is necessary, it is fit simultaneously to the purely thermal models since the determination of the shift is dependent on emission line features.
For the thermal with nonthermal model, the gain slope from the purely thermal model is used and held fixed while fitting the transient model.
Each FPM is given its own gain correction model, since the position of the emission within the detector FOV could affect the gain (e.g.~emission crossing the detector gap or near the edge of the FOV).

\begin{figure*}
  \begin{center}
    \includegraphics[width=\linewidth]{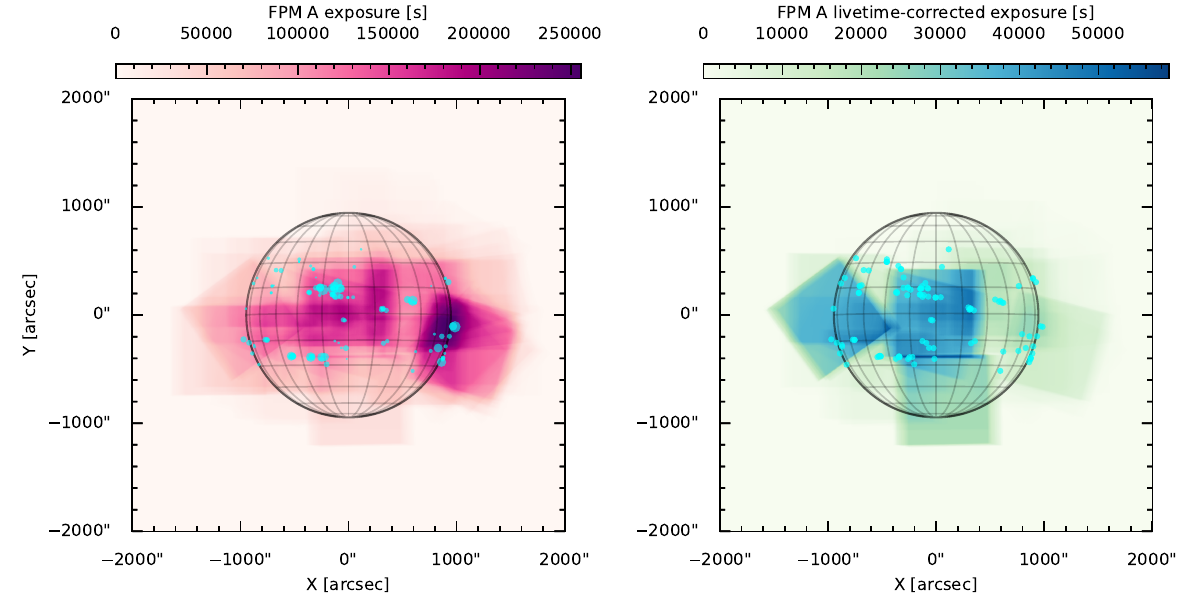}
  \end{center}
  \caption{Locations of all transients as seen by AIA (blue circles) overlaid on the wall-clock exposure time (\textit{left}) and the effective (i.e.~livetime-corrected) exposure time (\textit{right}) in FPM A. The circles denoting the transients in the left panel are scaled as sixteen times the AIA area while the circles in the right panel show their locations and are scaled equally. The total exposure time does not include the engineering test nor the mosaics.}%
  \label{fig:transient-locations-and-exposures}
\end{figure*}

\subsubsection{Thermal Energy Estimation}%
\label{subsubsec:thermal-energy-estimation}

The thermal energy content, $U_T$, of a transient is estimated with Equation~\ref{eq:thermal-energy},
\begin{equation}\label{eq:thermal-energy}
    U_T = 3 k_B T \sqrt{\text{EM} f V},
\end{equation}
using the temperature $T$ and emission measure $\text{EM}$ fit by the thermal models and volume $V$ from imaging.
The fill factor, $f$, is the fraction of the plasma volume emitting the observed photons, and $k_B$ is the Boltzmann constant.

A common practice, which is employed here, is to assume $f=1$, so our estimations then serve as upper limits on the thermal energy content.
\nustar's poor spatial resolution makes an automated measurement of $f$ in our study impossible with \nustar\ alone.
With the higher HXR resolution (down to \SI{2.3}{\arcsecond}) available with RHESSI,~\citet{Baylor-2011} followed up the RHESSI microflare statistical study (\citet{Christe-2008, Hannah-2008}, which assumed $f=1$) and computed $f$ for over \num{4500} microflares.
They found values as high as $10^{-1}$ and as low as $10^{-7}$, with a mean value of $10^{-3.7}$.
Their study demonstrates the variability in fill factors.

\begin{figure*}[t!]
  \centering
    \begin{minipage}[b]{.5\linewidth}
    \centering
    \includegraphics[width=\linewidth]{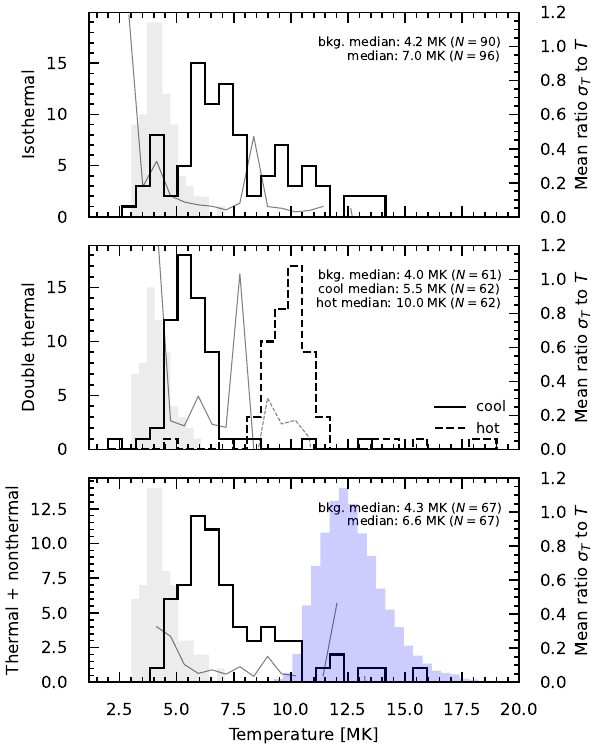}
    \end{minipage}\hfill
    \begin{minipage}[b]{.5\linewidth}
    \centering
    \includegraphics[width=\linewidth]{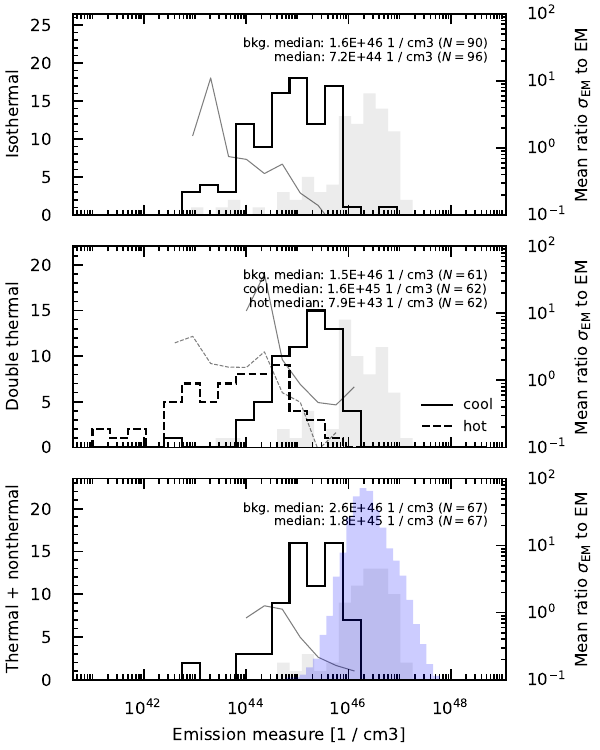}
  \end{minipage}
  \caption{Distributions of the thermal parameters for all three models and their background models. From top to bottom: isothermal, double thermal, and thermal with nonthermal. The gray-shaded histograms are the background (bkg.) model parameters, and the lines are the mean ratio of the uncertainties to their fit value for each histogram bin. The number of valid fits, $N$, for each histogram is listed next to their respective median. \nustar\ transients are cooler than RHESSI microflares but hotter than the quiescent corona. The RHESSI microflare data ($N=9161$) from~\citet{Hannah-2008} is shown by the shaded blue histograms in arbitrary units.}
  \label{fig:thermal-parameters}
\end{figure*}

EUV images from the Atmospheric Imaging Assembly (AIA) aboard the Solar Dynamics Observatory (SDO) were used to estimate plasma volumes ($V$) for our transients~\citep{Lemen-2012}.
We constructed proxy images of Fe \rom{18} emission using a linear combination of the 94, 171, and \SI{211}{\angstrom} filters~\citep{DelZanna-2013}.
\nustar\ has a strong response to temperatures that produce Fe \rom{18} emission, so these proxy images are commonly used to supplement \nustar\ analysis~\citep{Cooper-2020, Duncan-2021}.

We systematically examined each transient with level \num{1.5} AIA images, trimmed around a region containing the location of the transient obtained through \nustar.
Since the transients exhibit a broad range of temperatures (\SIrange{2.5}{15}{\mega\kelvin}; Figure~\ref{fig:thermal-parameters}), there was not one single AIA filter that could be used for all transients, so we iterated through various filters until it could be seen.
We began with the Fe \rom{18} proxy and subsequently searched through 131, 94, 171, then \SI{211}{\angstrom}, only proceeding to the next filter if the transient was not visible in the previous and stopping with the filter in which it was found.
The particular filter order was arbitrary except for the Fe \rom{18} starting point.
For a given filter, a map from near the X-ray emission peak was background subtracted using a map from the midpoint of the \nustar\ background interval, and we varied a threshold, filtering out lower-valued pixels, until the transient emission was isolated.
The area of the remaining pixels was used to compute the transient volume as $V=A^{3/2}$.
If it could not be isolated through thresholding the filters, a movie was used to estimate the area by examining plasma or loop motions; this was a common necessity for the fainter AR transients, where the plasma is subtly above the surrounding background temperatures.
This process was performed three times in order to set upper (maximum) and lower (minimum) limits on the areas, and the mean of the three trials was taken as the accepted value.

%% file: results.tex
\section{Results}%
\label{sec:results}

The automated identification and analysis methodology described in Section~\ref{sec:transient-identification-and-analysis} was applied to every \nustar\ solar observation from 2014 through 2022 (excluding the engineering test and mosaics), producing a catalog of transient candidates.
The results in this section feature the \num{113} true transients after removing false positives from pointing shifts, each categorized with one of the following labels: AR, QS, or ``intermediate regions'' (IR).
Classification as an AR transient required that the host region be a named NOAA active region at the time of the transient.
An IR transient was sourced from a region with increased magnetic complexity compared to the quiet Sun that would either become a named AR, was previously a named AR, or was never a named AR.
QS transients came from locations where there was no observed magnetic complexity at the time of the event.
The \num{113} transients were distributed as 62/35/16 (AR/IR/QS).
There are two transients for which AIA data does not exist, so no area was obtained for those and are thus excluded from any result involving volume or energy values.
We do not present any information regarding the temporal properties of our transients due to the time padding described in Section~\ref{subsubsec:grouping-clusters-in-time}.

\begin{table}[t!]
  \begin{tabular}{l c c c}
      Model & Transients & Background & Pileup \\
      \hline\hline
      T & 96 (50/32/14) & 90 & 52 \\
      T+T & 62 (34/24/4) & 61 & 37 \\
      T+N & 67 (43/24/0) & 67 & 47 \\
  \end{tabular}
  \caption{Number of valid fits for each transient model along with the number of AR/IR/QS transients shown in parentheses. The number of background and pileup corrections for each transient model are also listed.}%
  \label{tab:model-counts}
\end{table}

The locations of each transient on the solar disk is shown in Figure~\ref{fig:transient-locations-and-exposures} overlaid with exposures with FPM A.
Not all transients were triggered in both FPMs, likely due to our strict value for $q_d$, and a few transients had no valid fit models, whereby the fit to the data could not be adequately minimized.
This sample was further reduced by removing models that obviously performed poorly; namely, we filtered out the models that had parameters with fit values on the boundary of their limits.
We also removed any double thermal models that had two equivalent thermal components.
After filtering, we had \num{96} valid isothermal (T) models, \num{62} double thermal (T+T) models, and \num{67} thermal with nonthermal (T+N) models.
We do not choose a ``best'' model for any transient, so the same transient can appear in multiple categories.
Although, it is worth noting that prior work has used nested sampling to discriminate between multiple models by quantifying how well a given model represents the data relative to the others~\citep{Cooper-2024}.
Such an approach could be used in the future with our sample for identifying potential nonthermal emission.

See Table \ref{tab:model-counts} for a summary of how many fits required background and pileup corrections and the number of AR/IR/QS transients.
Most transients, including QS events, required a background component.
It can also be seen that there were no valid nonthermal fits for the QS transients, suggesting an isothermal model was often sufficient in representing the spectra.

\begin{figure}
  \begin{center}
    \includegraphics[width=\linewidth]{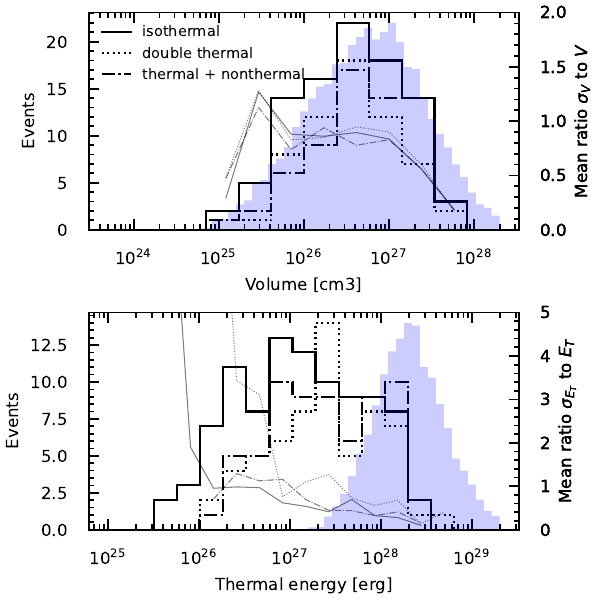}
  \end{center}
  \caption{Distributions of transient volumes (\textit{top}) and thermal energies (\textit{bottom}) for the transients with three different spectral models. The volume uncertainties are computed using the upper and lower area values derived as described in Section~\ref{subsubsec:thermal-energy-estimation}. The computed volumes are independent of the model but different subsets of transients are valid for the different models, hence the separate histograms. The thermal energies are computed from the \texttt{vapec} model parameters using Equation \ref{eq:thermal-energy}. The shaded blue histograms are from RHESSI~\citep{Hannah-2008}.}%
  \label{fig:volumes-and-thermal-energies}
\end{figure}

\subsection{Thermal Parameters}

\begin{figure*}
  \begin{center}
    \includegraphics[width=\linewidth]{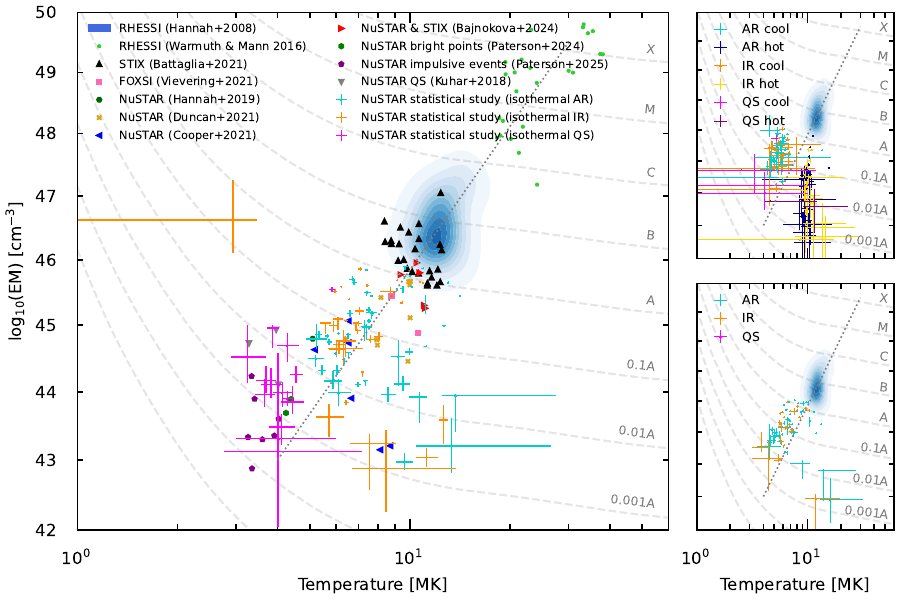}
  \end{center}%
  \caption{Emission measure versus temperature for several X-ray instruments across many magnitudes of GOES flare classes (indicated by the gray, dashed lines). The blue contours are the RHESSI microflares from~\citet{Hannah-2008}. The thermal parameters from the isothermal models (\textit{left}), double thermal models (\textit{top right}) split into cool and hot components, and thermal with nonthermal models (\textit{bottom right}) from this study are shown. The results are separated into AR, IR, and QS events and lie in trend with all other studies. For clarity, most instruments are omitted from the panels in the right column, leaving only the RHESSI contours for reference. The dotted gray line in all panels is purely a visual aid; it is not a fit line.}
  \label{fig:TEM-scatter}
\end{figure*}

The distribution of thermal parameters, $T$ and $\text{EM}$, are shown in Figure~\ref{fig:thermal-parameters} along with the background parameters respective to each model.
Since most transients were from ARs or IRs, the background models are predominantly characteristic of quiescent AR emission concentrated around \SIrange{3}{5}{\mega\kelvin}.
The transient temperatures span a wide range of \SIrange{2.5}{15}{\mega\kelvin}.
The errors on the temperatures are large at lower temperatures due to \nustar's lower sensitivity to cooler plasma while higher temperatures are better-constrained with errors lower than 20\%.
Our range of observed $\text{EM}$ values spans over three orders of magnitude, with some on the order of \SI{e42}{\per\centi\meter\cubed}, demonstrating \nustar's unmatched sensitivity.
However, uncertainties on $\text{EM}$ are considerably higher, with some on the same order of magnitude as the fit value itself.
Wide uncertainties on $\text{EM}$ values are common in more weakly energetic transients since they are more poorly constrained and more sensitive to perturbations in the parameter space~\citep{Duncan-2021,Cooper-2021}.
Thermal energies were computed for each transient from their model results and areas using Equation \ref{eq:thermal-energy}, with uncertainties computed using the lower and upper bounds of each fit parameter.

In comparison to the RHESSI microflares, the \nustar\ transients are generally cooler and dimmer, consistent with their thermal energy content being lower.
This is reflected in the bottom panel of Figure~\ref{fig:volumes-and-thermal-energies}, where we see that the derived thermal energies from all three of our transient models are comparable to each other and peak about an order of magnitude lower than RHESSI.
However, it can be seen (also in Figure~\ref{fig:volumes-and-thermal-energies}) that the volumes from the \nustar\ and RHESSI sets are similar, suggesting that the volume occupied by the emitting plasma could be independent of the energy released.

\begin{figure}
  \begin{center}
    \includegraphics[width=\linewidth]{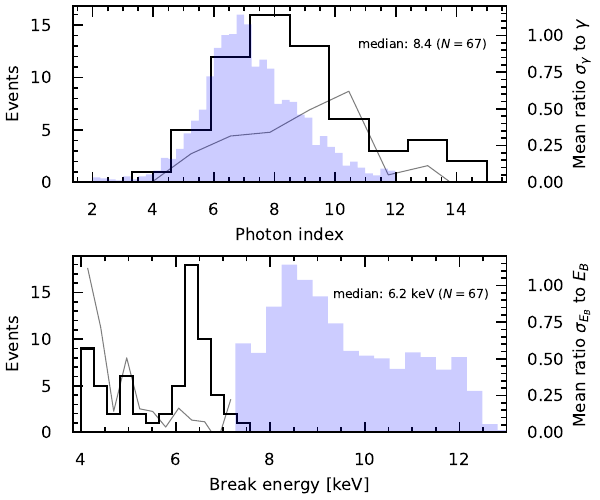}
  \end{center}
  \caption{Distributions of the nonthermal parameters. Like in Figure~\ref{fig:thermal-parameters}, the lines are the mean ratio of fit uncertainties for each bin, and the shaded blue histograms are from RHESSI~\citep{Hannah-2008}. The sharp left edge on the RHESSI break energy distribution is from a threshold set in the original analysis.}%
  \label{fig:nonthermal-parameters}
\end{figure}

\begin{figure}
  \begin{center}
    \includegraphics[width=\linewidth]{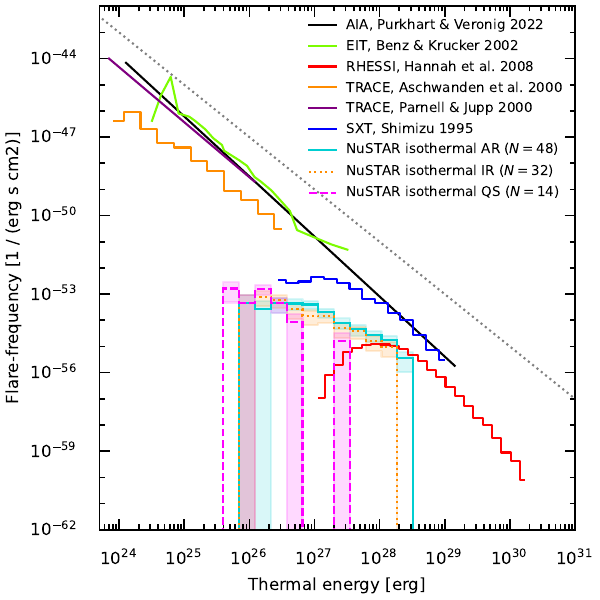}
  \end{center}
  \caption{Frequency distributions constructed from the \nustar\ isothermal model parameters divided among the three transient categories and plotted with distributions derived from other instruments. All three \nustar\ distributions were normalized by the same factor with their uncertainties obtained by normalizing the counting uncertainty by the area exposure. The RHESSI curve is the only other HXR distribution. Adapted from~\citet{Purkhart-2022} with measurements from this study added. The gray dotted line indicates a power law with a slope of \num{2}. The vertical location of each distribution can be subject to location within a solar cycle with a variability of several orders of magnitude.}%
  \label{fig:isothermal-frequency}
\end{figure}

\begin{figure}
  \begin{center}
    \includegraphics[width=\linewidth]{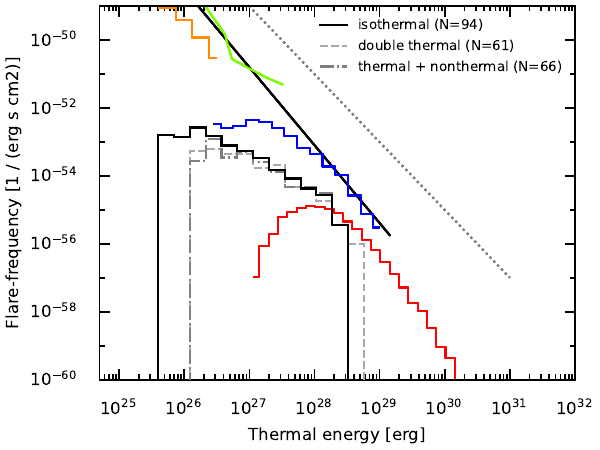}
  \end{center}
  \caption{Frequency distributions constructed from the results of our three spectral models. All models generally agree in shape and slope with the highest discrepancies found at lower energies, where the more complex models fail more often. The other distributions are the same as those shown in Figure~\ref{fig:isothermal-frequency} but left unlabeled for clarity.}%
  \label{fig:all-frequency-distributions}
\end{figure}

Figure~\ref{fig:TEM-scatter} presents the transients from this work alongside flares and transients observed by several other instruments, including \nustar\ microflares from prior studies.
Our isothermal results (separated by AR, IR, and QS events), match what has previously been published with \nustar\ and are in trend with other instruments.
The QS transients generally occupy a cooler and dimmer region of the parameter space compared to the AR and IR transients, implying the QS and AR+IR events may be two different populations (this is further explored in Section~\ref{subsec:ar-vs-qs-transients}).
Emphasis is placed on our isothermal results as that offers the most direct comparison to most of the other studies plotted since all, except for~\citet{Duncan-2021}, assume an isothermal approximation for the observed thermal emission.
The two RHESSI studies, one microflare from~\citet{Battaglia-2021}, one microflare from~\citet{Cooper-2021}, and three fits from~\citet{Bajnokova-2024} include a nonthermal component.
Our double thermal models (top right panel) partition the energy into two distinct groups, placing the hot component generally around \SI{10}{\mega\kelvin} and the cool component closer to AR temperatures.
These models typically display large uncertainties, particularly for lower GOES classes, indicating a two temperature model does not work well for fainter or less energetic transients seen by \nustar.
The thermal with nonthermal models (lower right panel) occupy a similar region of parameter space as the isothermal models but with fewer samples and at generally higher emission measures, indicating this model was unsuccessful for the fainter, less energetic transients.
It can be seen in Figure~\ref{fig:volumes-and-thermal-energies} that indeed there is a deficit of thermal with nonthermal fits at lower thermal energies compared to the isothermal fits.

There is an IR outlier that is exceptionally cool and bright, seen in Figure~\ref{fig:TEM-scatter}, that is a consequence of a very dynamic region with a particularly bright background.
On closer inspection, a thin strand of loops in the core of the region seems to rapidly heat and cool around the time X-rays were observed, suggesting it is a real transient.

\subsection{Nonthermal Parameters}

Figure~\ref{fig:nonthermal-parameters} shows the distributions of nonthermal parameters obtained from the \nustar\ fits, including RHESSI histograms for reference.
The photon indices measured by \nustar\ tend to be slightly steeper than those from RHESSI, indicating that the spectral hardness of microflares and fainter transients is not greatly different despite populating broadly different GOES classes (e.g.~Figure~\ref{fig:TEM-scatter}).
The measured break energies are impacted by instrumental and methodological biases; the RHESSI distribution was purposely limited to events with $E_B \geq \SI{7}{\kilo\eV}$ to remove events with ambiguity between the thermal/nonthermal emission.
The break energies observed by \nustar\ are much lower due to its good low-energy sensitivity.
However, many of the nonthermal fits selected a break energy near the Fe complex at \SI{\bettersim 6.7}{\kilo\eV}, suggesting the fit struggled to identify a ``true'' break in the spectrum.
While it is possible for the spectrum to break near the Fe complex, past \nustar\ analysis has indicated that this can indicate degeneracy in fit parameters.
Since a disproportionate number of fits have these values for $E_B$, we suspect that at least some of the fits are subject to this degeneracy.
We made no attempt to remove this since we fit a photon model rather than an electron model.
Investigation into this feature could be done more thoroughly in the future using an electron model (e.g. a thick- or thin-target model) and nested sampling.

\begin{figure}
  \begin{center}
    \includegraphics[width=\linewidth]{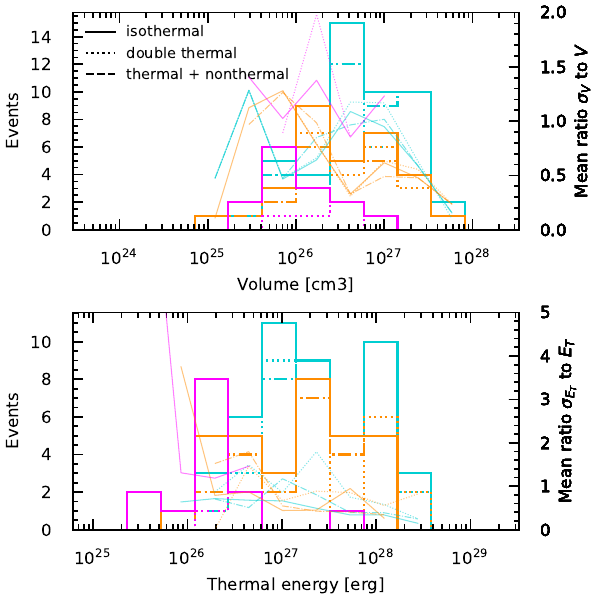}
  \end{center}
  \caption{Distributions of transient volumes and thermal energies separated into AR (blue), IR (orange), and QS (purple) by model. Like in Figure~\ref{fig:thermal-parameters}, the lines are the mean ratio of uncertainties for each bin. The volume uncertainties are computed using the upper and lower area values derived as described in Section~\ref{subsubsec:thermal-energy-estimation}.}%
  \label{fig:volumes-and-thermal-energies-ar-vs-qs}
\end{figure}

\subsection{Frequency Distributions}

Flare-frequency distributions were constructed from our model results.
The frequency distribution was normalized for exposure and visibility of the solar disk within \nustar's FOV, summed over both FPMs.
This normalization excludes the engineering test as well as any mosaic observations.
Since \nustar\ cannot observe the full solar disk, an area-time product, hereafter referred to as the area exposure, was computed for each observation.
This quantity accounts for \nustar's limited FOV, pointing drift, and exposure time by modeling \nustar's FOV as an array of polygons in time steps and is explained in detail in Appendix~\ref{app:frequency-normalization}.
Uncertainty is introduced into this quantity due to misalignment between \nustar's pointing solution and the actual location it observed on the Sun, but this uncertainty is small with respect to that from spectral fitting.
Our distributions are not corrected for the solar cycle (see Section~\ref{subsubsec:effect-of-the-solar-cycle}) nor for instrument or algorithm sensitivity.

The frequency distribution from the isothermal results as a function of thermal energy is shown in Figure~\ref{fig:isothermal-frequency}, separated by AR, IR, and QS events.
The three curves are normalized by the same factor rather than separately normalizing by ``active'' and ``quiet'' observations (see Section~\ref{subsec:ar-vs-qs-transients}).
Additionally, a comparison of the frequency distributions derived from each of our three spectral models is shown in Figure~\ref{fig:all-frequency-distributions}.
It can be seen that the \nustar\ distributions are slightly more shallow compared to RHESSI and SXT, but it is not clear whether this is a real effect or a consequence of either instrumental or algorithm sensitivity limitations imposed by our conservative criteria for identifying transients.
The distributions from the three \nustar\ models agree well with each other, except at the lowest energies due to the smaller number of weakly energetic events in our sample.
In particular, the QS events often failed for the double thermal and thermal with nonthermal models (Table~\ref{tab:model-counts}).

\begin{figure}
  \begin{center}
    \includegraphics[width=\linewidth]{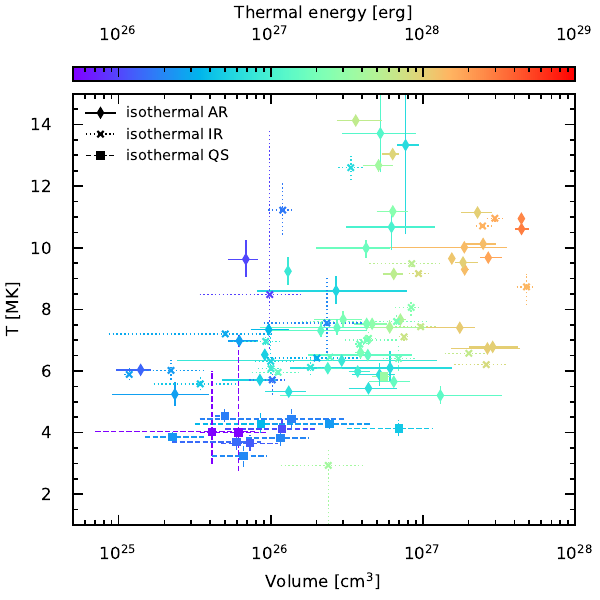}
  \end{center}
  \caption{Isothermal fit temperature versus thermal volume, separated into AR (thin diamonds with solid error bars), IR (``x'' with dotted error bars), and QS (squares with dashed error bars) events. The color indicates the derived thermal energy.}%
  \label{fig:temperature-vs-volume}
\end{figure}

%% file: discussion.tex
\section{Discussion}%
\label{sec:discussion}

We have presented a statistical collection of the faintest observable HXR transients to-date and distinguished them into three classes based on their host region: AR, IR, and QS.
In comparison to microflares, our observed transients (which include some traditional microflares) are cooler and fainter, and those with a valid broken power law fit suggest that the \nustar\ transients are slightly steeper compared to the microflares observed by RHESSI.
These characteristics are as expected for events with lower energy content.

\subsection{Comparison between AR and QS Transients}%
\label{subsec:ar-vs-qs-transients}

This is the first study to directly, systematically compare X-ray transients in ARs and the quiet Sun.
The IR transients behave similarly to the AR transients, so they will be considered part of the same group here.
Figure~\ref{fig:TEM-scatter} shows how the isothermal model results relate to prior studies for our AR and QS transients and indicates a separation in parameter space between the two, possibly due to different mechanisms driving their energy release.
The QS results from~\citet{Kuhar-2018} and flaring bright point results from~\citet{Paterson-2024} (also from \nustar, shown in the same figure) further support this.

We believe this feature of being cool but bright is a true characteristic of the QS transients, rather than instrumental, considering \nustar\ would be sufficiently capable of observing an event that is predominantly hot but faint given its temperature response (e.g.~\citet{Duncan-2024}).
Yet, the QS transients, from our study and those from~\citet{Kuhar-2018} and~\citet{Paterson-2024}, do not exhibit the same breadth of temperatures as those from ARs (Figure~\ref{fig:TEM-scatter}).
Thus far, \nustar\ has not provided evidence of impulsive events producing hot yet faint plasma in the quiet corona.
Better insight into the heating/cooling profiles of these events could be gained through differential emission measure (DEM) analyses, which may reveal a dim, hotter component by treating the plasma as a distribution rather than discrete temperatures.
For example,~\citet{Duncan-2024} identified the presence of \SI{12.6}{\mega\kelvin} plasma in time-evolved DEMs of a microflare in what a double thermal spectral fit represented as \SI{\bettersim 10}{\mega\kelvin}.
However, if such hotter plasma exists in these QS transients, it must be only in small quantities.
\citet{Paterson-2024} performed DEMs on two flaring bright points and confirmed no emission above \SI{5}{\mega\kelvin} in plasma that produced \SIrange{4}{4.5}{\mega\kelvin} temperatures from spectroscopy.

Comparing the thermal energies and volumes of the \nustar\ transients to those of RHESSI microflares implies that energy release \textit{within complex magnetic structures} could be independent of the volume occupied by the emitting plasma.
We found that the volumes of the AR transients seen by \nustar\ are similar to those seen in RHESSI microflares, despite their thermal energy content being considerably lower (Figures~\ref{fig:volumes-and-thermal-energies} and~\ref{fig:volumes-and-thermal-energies-ar-vs-qs}).
In contrast, the \num{14} QS transients are generally even cooler, fainter, and \textit{on average} occupy smaller volumes compared to the AR transients.
However, \textit{none} of the \nustar\ transients, including the bright points analyzed by~\citet{Paterson-2024}, occupy volumes that are smaller than any of the microflares seen by RHESSI.

It can be seen by comparing the AR distribution in Figure~\ref{fig:isothermal-frequency} to the isothermal distribution in Figure~\ref{fig:all-frequency-distributions} that the former begins to turn over around \SI{e27}{\erg} while the latter stays linear down to \SI{e26}{\erg} before turning over as it reaches the limits of instrument or algorithm sensitivity.
Fainter, or more weakly-energetic, AR transients are more difficult to detect due to the high AR background.
This suggests that the occurrence of transients follows the power law down to at least \SI{e26}{\erg} \textit{regardless of the magnetic complexity of its host region}, i.e. this is true for both AR- and QS-sourced transients.
However, we cannot make a firm statement regarding the energy of the weakest AR or IR transients, namely whether there is a cutoff or how much the AR frequency distribution truly turns over (if at all) at lower energies.
Due to sensitivity limitations of \nustar\ and given that no attempt was made to sensitivity-correct the distribution, it is unknown how many weakly energetic AR transients were missed.

The QS transients generally occupy smaller volumes and lower energies (Figure~\ref{fig:volumes-and-thermal-energies-ar-vs-qs}), and we observe a higher energy limit to QS transients as permitted by our sampling of the frequency distribution.
All but one of the QS transients had thermal energies below \SI{e27}{\erg}.
If it were common for QS transients to be hotter or even higher in energy, \nustar\ would have observed them given its high sensitivity to those temperatures, implying QS transients with such high temperatures or energies either do not occur at all or are exceedingly rare.
Further, Figure~\ref{fig:temperature-vs-volume} exhibits a slight correlation between $T$ and $V$ and a separation between AR+IR and QS populations, with all but one of the QS transients cooler than \SI{5}{\mega\kelvin}.
This provides additional, independent evidence that the QS transients are a distinct population given that $T$ and $V$ were obtained through completely separate procedures.
All nonthermal fits for QS transients failed, so we cannot make any statement about their properties.

Figure~\ref{fig:isothermal-frequency} shows that the AR, IR, and QS distributions all reach similarly low energies (\SI{\bettersim e26}{\erg}), while the QS distribution does not extend beyond \SI{e27}{\erg} aside from an outlier around \SI{2e27}{\erg}.
It should also be noted that the relative magnitudes of the AR+IR and QS frequency distributions should not be compared, as all curves were normalized by the same factor.
We did this because \nustar's dynamic range is not characterized as a function of its livetime (see Section~\ref{subsec:observational-biases}), so it is unknown how weakly energetic an observed QS transient can be while a brighter structure, such as an AR, is within the FOV.

\subsection{Instrumental Discrepancies in the Frequency Distribution}

We observed transients spanning nearly three decades in energy with \nustar.
The \nustar\ frequency distribution (Figure~\ref{fig:isothermal-frequency}) overlaps with RHESSI, beginning just where the RHESSI distribution starts to turn over due to sensitivity limitations.
While the \nustar\ and RHESSI distributions are well-aligned, there is an apparent shift between them and all others.
This discrepancy is likely a consequence of biases introduced by HXR-only or EUV-only analyses, and here we will discuss the factors governing the relative position of the distributions shown in Figure~\ref{fig:isothermal-frequency}.

The frequency (vertical scaling) of AR-sourced events is governed by solar activity and location within the solar cycle (\nustar\ is an exception due to observational biases; see Section~\ref{subsubsec:effect-of-the-solar-cycle}).
Nearly an order of magnitude variability was observed in RHESSI microflares over just four years~\citep{Hannah-2008}.
Conversely,~\citet{Purkhart-2022} found that QS EUV nanoflares are independent of the solar cycle, and this independence may explain why the distributions from the other EUV nanoflare analyses shown in Figure~\ref{fig:isothermal-frequency} generally agree well despite utilizing a much lower volume of data (on the order of tens of minutes).

The energetics (horizontal scaling) of the distributions are heavily subject to methodological biases.
Narrowband EUV filters, such as TRACE and EIT, will systematically underestimate energies of hotter nanoflares, resulting in an artificially steeper and energetically narrow distribution~\citep{Aschwanden-2002}.
AIA consists of a suite of filters, thus enabling the use of DEM analysis; however, it is subject to a ``blind spot'' in $T$-$\text{EM}$ space, which is a well-known characteristic of EUV and non-spectrally-resolved SXR instruments~\citep{Winebarger-2012,Athiray-2024}.
In DEM analyses, the blind spot often manifests as an overabundance of hot plasma---and thus an overestimation of thermal energy content---unless HXR~\citep{Duncan-2024} or spectrally resolved SXR data~\citep{Athiray-2024} are used to constrain the parameter space.
A survey by~\citet{Warmuth-2020} examined five studies from the past two decades (some including RHESSI and AIA) and concluded that the largest uncertainty in the thermal energy of a hot plasma is the determination of the DEM.

Both \nustar\ and RHESSI rely on HXR spectroscopy, which will always characterize the hottest plasma.
It has been shown by \citet{Inglis-2014} that, for more energetic events such as microflares, the hottest plasma contributes most of the thermal energy.
Their analysis revealed that a DEM constrained by both RHESSI and AIA produced similar energies as a RHESSI-only isothermal fit.
However, it is unknown if this remains true for weaker, cooler events, such as our QS transients, where cold ($<\SI{2}{\mega\kelvin}$) plasma could contribute a higher fraction to the total thermal energy content of the event.
It is possible that HXR spectroscopy could begin to underestimate energies for QS transients.
Additionally, the assumption of a filling factor of $f=1$ (Section~\ref{subsubsec:thermal-energy-estimation}) produces an upper limit on thermal energies, while it has been shown that $f$ spans several orders of magnitude~\citep{Baylor-2011}.
A different value of $f$ would only shift the HXR distributions to \textit{lower} energies and thus maintains our observed lack of higher energy (\SI{>e27}{\erg}) QS transients.

Despite these biases in energy, we can conclude that there is a distinct lack of high energy, hot QS transients.
\citet{Purkhart-2022} found EUV events in the quiet Sun with energies up two orders of magnitude higher than what we found with \nustar.
Those high energy EUV events \textit{must} be cold and outside of \nustar's temperature sensitivity, otherwise \nustar\ would have observed similarly high energy events in the quiet Sun.

\subsection{\nustar's Observational Biases}%
\label{subsec:observational-biases}

\nustar's limited photon throughput of \SI{400}{\count\per\second} per telescope limits observation of events more energetic than $\bettersim$high-A-class microflares because it prohibits detection of photons greater than \SIrange{\bettersim 10}{12}{\kilo\eV}; the combination of the limited spectral dynamic range and the steeply falling solar spectra causes the sharp drop on the right side of the $E_B$ distribution in Figure~\ref{fig:nonthermal-parameters}.
We emphasize that care must be taken with detection claims of nonthermal emission for weakly energetic transients (e.g.~sub-A-class microflares).
Several studies have produced strong evidence of nonthermal emission in \nustar\ spectra of faint microflares through meticulous analysis, including a GOES A0.02-class equivalent~\citep{Glesener-2020,Cooper-2021,Cooper-2024,Bajnokova-2024}.

Additionally, \nustar's limited throughput requires that the brightest structures on the solar disk be within its FOV, lest the instrument be flooded with ghost rays~\citep{Grefenstette-2016}.
This makes it difficult to perform QS observations (except at solar minimum), resulting in most observations containing bright structures (e.g. ARs).
A major methodological limitation in our study was the use of the moving median for background estimation, prohibiting us from searching mosaic observations.
In total, there are about \SI{13}{\hour} (wall clock time) of mosaic observations, with most of this time during relatively quiet intervals.
Consequentially, only about a tenth of the transients in our catalog are QS despite the expectation that less energetic events should be much more frequent (e.g.~Figure~\ref{fig:isothermal-frequency}).

Another effect of the limited throughput is the time interval over which we performed spectroscopy.
In contrast to the \SI{16}{\second} intervals at the peak used for the RHESSI microflare analysis~\citep{Christe-2008,Hannah-2008}, we imposed a minimum of \SI{2}{\minute} for our event durations, often including portions of the impulsive and decay phases of the transient.
Inclusion of the decay phase likely results in a higher emission measure (and lower temperature) and thus a different estimated thermal energy compared to an impulsive- or peak-only time interval~\citep{Glesener-2017}.

Due to \nustar's large HPD, we had to use EUV images from AIA in order to estimate our plasma volumes.
This, combined with our assumption of a fill factor of unity, could introduce a bias on our derived volumes (hence, thermal energies), likely through overestimation.
\citet{Kuhar-2018} claims an overestimate up to a factor of \num{5} is possible in their analysis of three QS transients due to the broader temperature response from the EUV filters compared to \nustar.
Given these differences, our volumes and thermal energies can be treated as upper limits on the observed transients.
Determination of any systematic differences between the EUV-derived volumes and HXR-derived volumes could be performed by examining a set of microflares co-observed by RHESSI and AIA.

\begin{figure}
  \begin{center}
    \includegraphics[width=\linewidth]{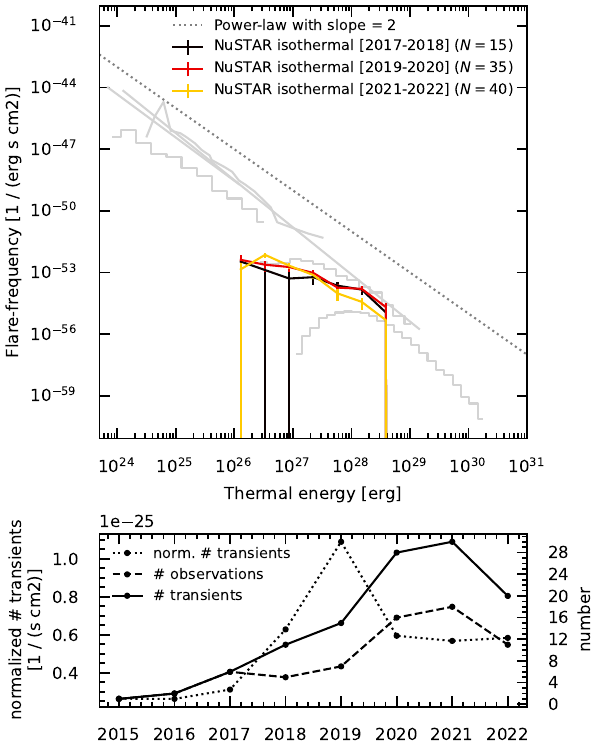}
  \end{center}
  \caption{\textit{Top}: frequency distributions constructed using our results chunked in three two year-long intervals. Each distribution is normalized according to the observations during the corresponding time interval. The errorbars are from counting statistics. The lightgray lines are the same distributions from other instruments shown in Figure~\ref{fig:isothermal-frequency}. \textit{Bottom}: number of transients (solid) and observations (dashed) each year analyzed in our study (right-hand axis). The dotted line shows the number of transients normalized by the area exposure quantity for each year (left-hand axis).}%
  \label{fig:solar-cycle-search}
\end{figure}

\subsubsection{Effect of the Solar Cycle}%
\label{subsubsec:effect-of-the-solar-cycle}

The occurrence rate of microflares is correlated with the solar cycle while fainter transients or ``nanoflare-like'' events are not~\citep{Purkhart-2022}.
For microflares, the distribution normalization, rather than the slope, is cycle-dependent~\citep{Aschwanden-2002, Hannah-2011}.
This effect can be seen in the RHESSI-observed microflares of~\citet{Hannah-2008}, where the shape is unchanged while the normalization varies by nearly an order of magnitude across four years.

We measure no significant change in transient frequency with solar cycle, although note that \nustar\ poorly samples the cycle.
The top panel in Figure~\ref{fig:solar-cycle-search} shows three frequency distributions chunked at two year-long intervals: \numrange{2017}{2018}, \numrange{2019}{2020}, and \numrange{2021}{2022} (\numrange{2015}{2016} was excluded since we only found two transients during that interval).
If the \nustar\ distributions were affected by the solar cycle, we would see a vertical shift between at least two of the lines shown, but we instead see an overlap of all three.
Indeed, the \numrange{2019}{2020} interval is the closest to solar minimum but is not below the other two lines.
This is corroborated by the bottom panel, which shows the number of observations and transients per year, as well as the number of transients normalized by the area exposure for the given year.
It can be seen that the number of transients observed per year is strongly correlated with the number of observations performed that year.
The normalized line is also correlated with the number of observations except between \num{2019} and \num{2020}.
In contrast,~\citet{Hannah-2008} found that the number of microflares observed per year varies directly with the solar cycle (e.g.~progressing into solar minimum results in fewer microflares), as microflares are sourced from active regions.
Since most of the \nustar\ transients in this study are from ARs, we expect to see a \textit{decrease} in number of transients per year as we progress towards \numrange{2019}{2020}.
Instead, the observed correlation implies that any effect from the solar cycle on \nustar\ transient frequency is dominated by observational biases.

%% file: conclusion.tex
\section{Conclusions}%
\label{sec:conclusions}

We presented the first statistical study of \nustar\ solar transients, comprehensively extending the sample of faint HXR transients down to \SI{e26}{\erg}.
Events were identified within active regions and the quiet corona, allowing us to directly compare the properties of each.
All of our quiet Sun transients are cool, all below \SI{5}{\mega\kelvin} but bright, matching what has previously been found in very faint quiescent emission observed by \nustar.
This population of transients appears to be distinct from the population of transients from more magnetically complex structures (e.g. active regions).

Using the \nustar\ catalog, we were able to characterize weakly energetic transients observed in HXRs down orders of magnitude beyond what was established with RHESSI.
Compared to RHESSI microflares, the \nustar\ transients exhibit spectra that are generally cooler, dimmer, and steeper, which is a consequence of their lower energy content.
Plasma associated with transients produced by active regions (as observed by \nustar\ or RHESSI) generally occupies similar volumes across all observed energies.
In contrast, plasma volumes in \nustar\ quiet-Sun transients tend to be smaller.

\nustar, while not optimized for solar observation, has demonstrated great value in measuring faint solar transients, including microflares.
However, its shortcomings include limited spatial resolution and photon throughput, resulting in our catalog being heavily biased towards active region transients.
This stresses the need for a solar-dedicated, HXR telescope that constantly observes the Sun and is capable of detecting faint transient emission.

%% file: acknowledgements.tex
\begin{acknowledgements}
    Reed B. Masek's research is supported under the NASA FINESST 2022 award (80NSSC23K1621).
    Sarah Paterson and Nat{\'a}lia Bajnokov{\'a} acknowledge support from the UK's Science and Technology Facilities Council (STFC) doctoral training grants (ST/T506102/1 and ST/X508391/1).
    Iain G. Hannah acknowledges support from STFC grants (ST/T000422/1, ST/X000990/1).
    This research has made use of data and/or software provided by the High Energy Astrophysics Science Archive Research Center (HEASARC), which is a service of the Astrophysics Science Division at NASA/GSFC.
    This research has made use of data from the \nustar\ mission, a project led by the California Institute of Technology, managed by the Jet Propulsion Laboratory, and funded by the National Aeronautics and Space Administration. Data analysis was performed using the \nustar\ Data Analysis Software (NuSTARDAS), jointly developed by the ASI Science Data Center (SSDC, Italy) and the California Institute of Technology (USA).
    All \nustar\ observations were completed as part of the NASA \nustar\ Guest Observer program.
    \textit{Software:} aiapy~\citep{Barnes-2020-aiapy}, sunpy~\citep{Barnes-2020-sunpy}, astropy~\citep{Price-2022-astropy}, scipy~\citep{Virtanen-2020-scipy}, numpy~\citep{Harris-2020-numpy}, matplotlib~\citep{Hunter-2007-matplotlib}.
\end{acknowledgements}

%% file: detector-gap-examination.tex
\section{Detector Gap Examination}%
\label{app:detector-gap-examination}

In Section~\ref{subsubsec:region-selection}, we outlined our procedure for selecting a region for spectroscopy and defined the ``detector gap'' formed by the four closely-spaced CdZnTe detectors that make up an individual FPM.
\nustar\ solar spectroscopy typically avoids crossing this gap due to the known greater uncertainty in the detector response near the edges of each sensor.
We urge that care must still be taken with transients around the detector gap when examining individual events since results \textit{do} change on a case-by-case basis.

\begin{figure*}
  \centering
  \includegraphics[width=\linewidth]{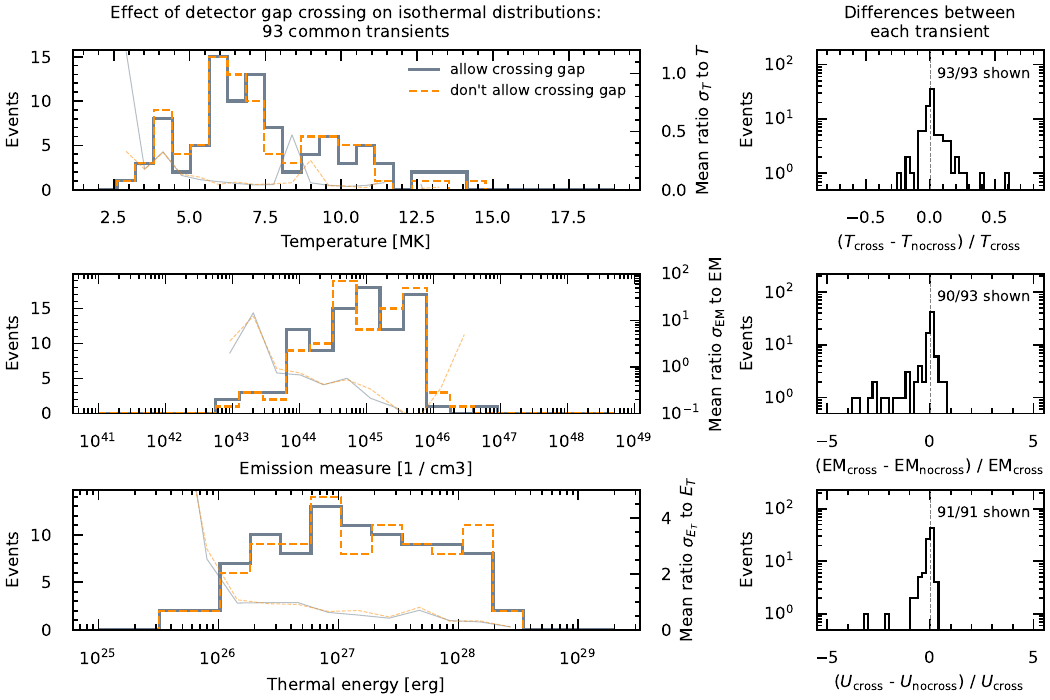}
  \caption{Comparison of the isothermal distributions when allowing and disallowing the spectral regions to cross the detector gap. The left column shows the temperature, emission measure, and thermal energy distributions along with their errors, like shown in previous figures. The right column shows the differences between the parameters for each transient. The numbers at the top right of the plots in the right column denote the number of transients represented in the distributions. There were a few outliers in the emission measure distribution, indicating those were significantly affected by the detector gap condition. Recall that there are two transients for which volumes (thus, energies) could not be derived, hence there are \num{91} events shown in the thermal energy distributions rather than \num{93}.}
  \label{fig:detectorgap-effect}
\end{figure*}

Figure~\ref{fig:detectorgap-effect} compares the isothermal distributions with and without the gap avoidance condition.
There were \num{93} transients common to both datasets, meaning \num{3} transients that produced valid isothermal fits when allowing the regions to cross the detector gap did \textit{not} produce valid fits when the regions were not allowed to cross the gap.
The distributions in the right column show the relative change in fit values for each transient, with most changes clustered around zero.
As denoted by the numbers in those plots, there were a few transients that lie outside the range shown on the plots, indicating they were greatly affected by the difference caused by the detector gap condition.
Indeed, the three transients outside the indicated emission measure range had the transient and/or the background region on or near the gap, making the spectral fits particularly sensitive to the region selection.

An important caveat is that we do not have absolute certainty on the location of the detector gap within the FOV.
\nustar\ data is returned as an event list that has the spatial and temporal properties of each photon that was detected, including which detector quadrant recorded the photon.
We estimated the location of the gap by requiring that the photons within the proposed spectroscopy region be associated with only \textit{one} detector quadrant.
If, in order to enforce this requirement, the radius of the region shrunk below \SI{25}{\arcsecond}, we set the radius to equal \SI{25}{\arcsecond}.
Although this minimum value is already quite small (considering \nustar's HPD is about \SI{60}{\arcsecond}), this was to prevent the region from being \textit{too} small and thus excluding transient-souced photons, even if it meant it had to slightly cross the gap.

We allowed our regions to cross this gap after we did this side-by-side comparison of the spectroscopy results for our transients and found that the \textit{bulk} behavior of our results did not change, so the conclusions of our analysis are unaffected.
We chose to present our results based on the dataset with the regions that were allowed to cross the gap so that the emission from our transients was not trimmed by our automated region selection since \num{85} out of the \num{113} transients had at least one FPM that was affected by the detector gap.

%% file: frequency-normalization.tex
\newpage
\section{Frequency Normalization}%
\label{app:frequency-normalization}

\nustar\ cannot image the full solar disk, so we numerically computed contributions to the frequency distribution normalization for each observation and summed them together.
Exposure images were made at fixed intervals of \SI{5}{\minute} from the start to the end of the observation.
For each image, \nustar's FOV was estimated (as described in Section~\ref{subsubsec:region-selection}) and a \numproduct{3000 x 3000} grid of points was drawn within.
This two-dimensional grid was then projected onto a three-dimensional sphere representing the photosphere.
Polygons were drawn by connecting three or four neighboring points within the projected grid, and the sum of all polygon areas was taken to be the estimate of the amount of solar surface area within the FOV.
The areas and their respective exposure times were multiplied together and tracked for all observations, and the sum of values from all observations was used as the normalization factor for the frequency distribution.
We removed the first and last five minutes' worth of data from the calculation for each observation to account for the loss of time introduced by the moving median.
Lastly, our calculation does \textit{not} remove area ``lost'' to the detector gap since it is suspected that \nustar\ would be able to detect an event even if it were positioned exactly in the middle of the gap due to \nustar's large point spread function.

\begin{figure*}
  \centering
  \includegraphics[width=0.8\linewidth]{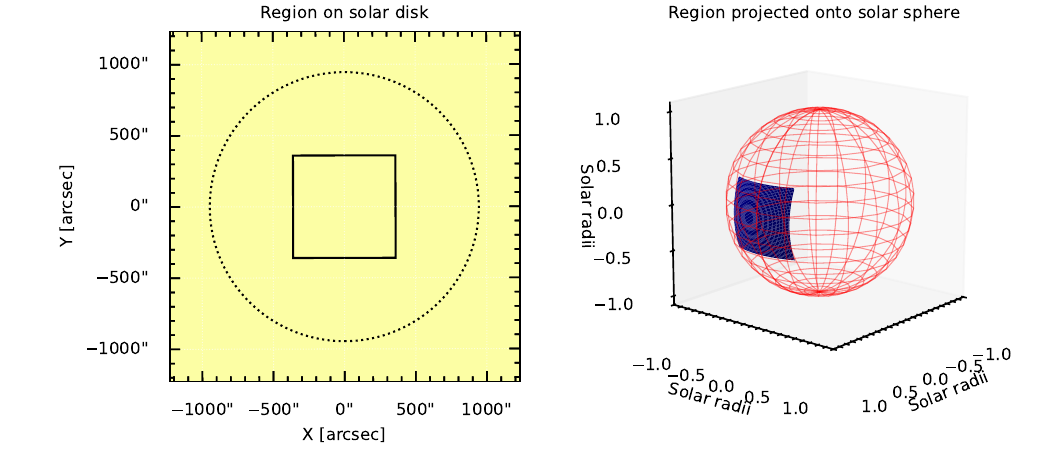}
  \includegraphics[width=0.8\linewidth]{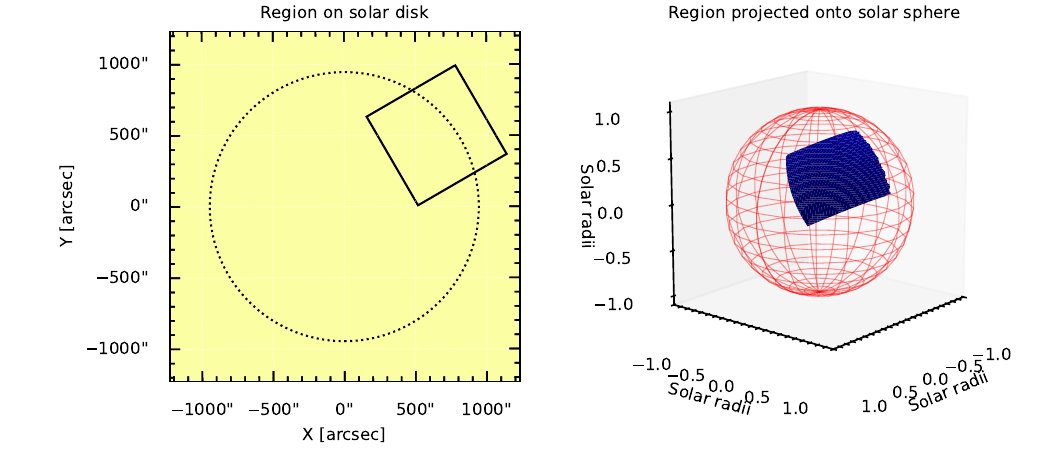}
  \caption{\textit{Top row}: example of a black square representing \nustar's FOV (\textit{left}) projected on a sphere (\textit{right}). \textit{Bottom row}: same as the top but for a FOV on the limb of the disk, demonstrating the projection effect. Example of a black square representing \nustar's FOV (\textit{left}) projected onto a sphere (\textit{right}).}
  \label{fig:centered-fov}
\end{figure*}

This method consistently underestimates the true area observed by \nustar\ since the individual polygons are flat and do not curve along the spherical surface.
However, we know that this error is approximately 1\% at the grid resolution used in our calculations.
This error in the normalization is small in comparison to other errors in our frequency distribution, such as those associated with the spectral fits.
Our normalization does not include the possibility of detecting occulted events given that the detection of occulted events is highly dependent on the loop geometry and altitude.

%% file: table.tex
\newpage

\section{Table of Spectral Fit Parameters}%

\begin{deluxetable*}{c c c}
    \tabletypesize{\footnotesize}
    \tablewidth{0pt}
    \tablecaption{Spectral fit symbol definitions \label{tab:table-definitions}}
    \tablehead{\colhead{Symbol} & \colhead{Unit} & \colhead{Description}}
    \startdata 
    $T$ & \si{\mega\kelvin} & \texttt{vapec} temperature component \\
    $\text{EM}$ & \SI{e44}{\per\centi\meter\cubed} & \texttt{vapec} emission measure component \\
    $N$ & \si{\photon\per\keV\centi\meter\squared\second} & \texttt{bknpower} normalization at \SI{1}{\kilo\eV} \\
    $\gamma$ &  & \texttt{bknpower} high energy slope \\
    $E_B$ & \si{\kilo\eV} & \texttt{bknpower} break energy \\
    $F$ &  & factor between the two FPMs \\
    $G$ &  & gain correction to the response \\
    \hline
    $\lambda$ & \si{\count\per\kilo\eV\per\second} & exponential rate in pileup model \\
    $\mu$ & \si{\kilo\eV} & the Gaussian mean in pileup model \\
    $\sigma$ & \si{\kilo\eV} & the Gaussian standard deviation in pileup model \\
    $N_{A,B}$ &  & normalization in pileup model (FPM A or B) \\
    \hline
    $C$ &  & model fit statistic \\
    $\nu$ &  & model degrees of freedom (number bins minus number parameters) \\
    \hline
    $V$ & \SI{e25}{\centi\meter\cubed} & thermal volume from AIA \\
    $U_T$ & \SI{e26}{\erg} & thermal energy from fit parameters \\
    \enddata
    \tablecomments{Definitions for the symbols used in the spectral fit table, provided units and brief descriptions.}
\end{deluxetable*}


\begingroup
\everymath{\scriptscriptstyle}
\addtolength{\tabcolsep}{-0.5em}
\begin{longrotatetable}
    \movetabledown=0.7in
    \noindent
    \begin{longtable*}[]{c || cc ccc || cc ccc | ccc ccc | ccc ccc}
    \toprule
    \multirowcell{8}[0ex][c]{Date\\ID\\Bkg. time\\Event time} & \multicolumn{5}{c||}{\textbf{Background}} & \multicolumn{17}{c}{\textbf{Transient}} \\
    & & & & & & & \\
                                       & \multicolumn{5}{c||}{\small Isothermal} & \multicolumn{5}{c|}{\small Isothermal} & \multicolumn{6}{c|}{\small Double thermal} & \multicolumn{6}{c}{\small Thermal with nonthermal} \\
                                       & \multicolumn{2}{c}{\small\texttt{vapec}} & \multicolumn{3}{c||}{\small\texttt{pileup}} & \multicolumn{2}{c}{\small\texttt{vapec}} & \multicolumn{3}{c|}{\small\texttt{pileup}} & \multicolumn{3}{c}{\small\texttt{vapec}$+$\texttt{vapec}} & \multicolumn{3}{c|}{\small\texttt{pileup}} & \multicolumn{3}{c}{\small\texttt{vapec}$+$\texttt{bknpower}} & \multicolumn{3}{c}{\small\texttt{pileup}} \\

    \cmidrule(lr){2-3} \cmidrule(lr){4-6} \cmidrule(lr){7-8} \cmidrule(lr){9-11} \cmidrule(lr){12-14} \cmidrule(lr){15-17} \cmidrule(lr){18-20} \cmidrule(lr){21-23}

     & $T$ & $C$ & $\lambda_A$ & $\lambda_B$ & $C_A$ & $T$ & $C$ & $\lambda_A$ & $\lambda_B$ & $C_A$ & $T_1$ & $T_2$ & $C$ & $\lambda_A$ & $\lambda_B$ & $C_A$ & $T$ & $N$ & $C$ & $\lambda_A$ & $\lambda_B$ & $C_A$ \\ 
     & $\text{\fontsize{6pt}{9pt}\selectfont EM}$ & $\nu$ & $\mu_A$ & $\mu_B$ & $\nu_A$ & $\text{\fontsize{6pt}{9pt}\selectfont EM}$ & $\nu$ & $\mu_A$ & $\mu_B$ & $\nu_A$& $\text{\fontsize{6pt}{9pt}\selectfont EM}_1$ & $\text{\fontsize{6pt}{9pt}\selectfont EM}_2$ & $\nu$ & $\mu_A$ & $\mu_B$ & $\nu_A$ & $\text{\fontsize{6pt}{9pt}\selectfont EM}$ & $\gamma$ & $\nu$ & $\mu_A$ & $\mu_B$ & $\nu_A$ \\
     & $F$ & --- & $\sigma_A$ & $\sigma_B$ & $C_B$ & $F$ & $V$ & $\sigma_A$ & $\sigma_B$ & $C_B$ & $F$ & --- & $V$ & $\sigma_A$ & $\sigma_B$ & $C_B$ & $F$ & $E_B$ & $V$ & $\sigma_A$ & $\sigma_B$ & $C_B$ \\
\end{longtable*}
\end{longrotatetable}
\endgroup